\documentclass{article}
\usepackage{spconf, amsmath, graphicx, color, subcaption, booktabs, amssymb, url}
\usepackage{cite}

\title{Speech-to-Singing Conversion in an Encoder-Decoder framework}

\name{Jayneel Parekh$^{\star \dagger}$ \qquad Preeti Rao$^{\dagger}$ \qquad Yi-Hsuan Yang$^{\mathsection}$ }
\address{$^{\star}$ LTCI, T{\'e}l{\'e}com Paris, Institut Polytechnique de Paris, France\\
$^{\dagger}$ Department of Electrical Engineering, Indian Institute of Technology Bombay, India\\
$^{\mathsection}$ Research Center for IT Innovation, Academia Sinica, Taiwan\\}
\begin{document}
\ninept
\maketitle
\begin{abstract}
In this paper our goal is to convert a set of spoken lines into sung ones. Unlike previous signal processing based methods, we take a learning based approach to the problem. This allows us to automatically model various aspects of this transformation, thus overcoming dependence on specific inputs such as high quality singing templates or phoneme-score synchronization information. Specifically, we propose an encoder--decoder framework for our task. Given time-frequency representations of speech and a target melody contour, we learn encodings that enable us to synthesize singing that preserves the linguistic content and timbre of the speaker while adhering to the target melody. We also propose a multi-task learning based objective to improve lyric intelligibility. We present a quantitative and qualitative analysis of our framework.
\end{abstract}

\section{Introduction}

How often have we wanted to sing a few lines in tune? How often have music composers wished to hear multiple melodic variations of the same lyrical phrase? Building systems to cater to these goals would not only result in interesting applications but also enable us to explore relevant research questions in the audio domain. In this work, we propose a speech-to-singing (STS) system that attempts to solve the following problem: Given an input speech waveform and a target melody contour, transform the speech into a singing waveform that follows the target melody while preserving the speaker identity and linguistic content.

Our task of transforming an audio in a manner which preserves some of its characteristics (e.g., speaker identity, linguistic content), while modifying certain others (melody, phoneme durations) can be seen as an instance of the general problem of style transfer \cite{alexeyss}. There have been several recent works which can be viewed in a similar light, that is, they intend to transform only specific properties of audio (referred to as ``style'') while keeping others intact (referred to as ``content''). Conditioned on the input and the target timbre, Haque \emph{et al}. \cite{pverma1} propose a system to transform the timbre of the speaker/instrument while preserving the words/notes. Mor \emph{et al}. \cite{music_translation} propose to translate across various musical instruments, styles and genres. They use a single encoder to process all types of audio, and one decoder each to model every style/instrument/genre etc. Wu \emph{et al}. \cite{eric_sing} propose a Cycle-GAN \cite{cgan} based framework to perform singing voice conversion between any two singers and validate their results by performing gender transformation for singing voices.
``Style" assumes a very different meaning for our problem compared to other works on style transfer, as it not only refers to incorporating the desired melody but also requires to modify the speaking timbre to a singing timbre while preserving the speakers identity. There is another crucial aspect of ``style'' unique to our problem which is that of modelling phoneme durations in the predicted singing, which can be fairly different from the phoneme durations in speech.

As discussed in \cite{survey_sp2si}, even though speech and singing signals are produced by same vocal production system and consequently share many properties, their production takes place within very different settings. The key challenges for achieving conversion involve aligning two very different signals (equivalent to modelling phoneme durations), imposing the required melody on speech without losing linguistic content and speaker identity. Previous works on STS conversion can be broadly categorized into two approaches \cite{survey_sp2si}:

\textit{Model-based STS}: Work by Saitou \emph{et al}. \cite{gotosp2si} is the representative for these type of methods. Apart from the input speech and musical score they also take as input the phone-score synchronization information. This refers to the manual segmentation of the phonemes, and associating each with the corresponding musical note. They analyze the input speech and modify $F_0$, phoneme durations and spectral envelope using synchronization information and various manually designed control models. The modified components are used to synthesize the output singing.

\textit{Template-based STS}: Originally proposed in \cite{tsts1}, this class of methods assume that a high  quality-sung vocal (referred to as ``template singing'') is  available as a reference. The input speech and template singing are first aligned with each other. This itself is an active field of research with several proposed alignment approaches \cite{tsts1, dualalign, align_intsp} which often utilize features extracted by analyzing the speech/singing spectra. The template singing is further used to extract the reference prosody which includes singing $F_0$, aperiodicity index, singing formants, etc. This information is used to estimate parameters for singing synthesis from aligned speech. Concurrent work by Gao \emph{et al}. \cite{sts_spectral} (published in November 2019) proposes a deep learning based system for template-based STS conversion which conditions the network on $i$-vector of the speaker while predicting the singing spectral parameters to preserve speaker identity. However, as discussed below, all the aforementioned works significantly differ from our approach.

Prior arts are limited in their applicability as they require additional inputs like high quality singing templates \cite{sts_spectral, tsts1} or phone-score synchronisation information \cite{gotosp2si}, which can be very difficult to deduce for any general speech and melody. We thus opt for an approach to employ minimal additional information over the melody contour to achieve STS conversion. To our best knowledge, we are the first to attempt such a transformation using a machine learning based method that does not require any phoneme synchronization information or high-quality singing template. We do so in an encoder-decoder framework and propose a novel multi-task learning (MTL) based enhancement for improving phoneme intelligibility.

We are interested in investigating the performance of supervised approaches and assume availability of paired speech-singing examples for our task. While it is interesting to explore unsupervised methods for this task, supervised style transfer approaches are known to outperform unsupervised methods \cite{cgan}. To this end, We utilize the largest public paired dataset for STS conversion, the NUS sung and spoken lyrics corpus \cite{nus48e}.

\begin{figure}
    \centering
    \includegraphics[width=1.0\columnwidth]{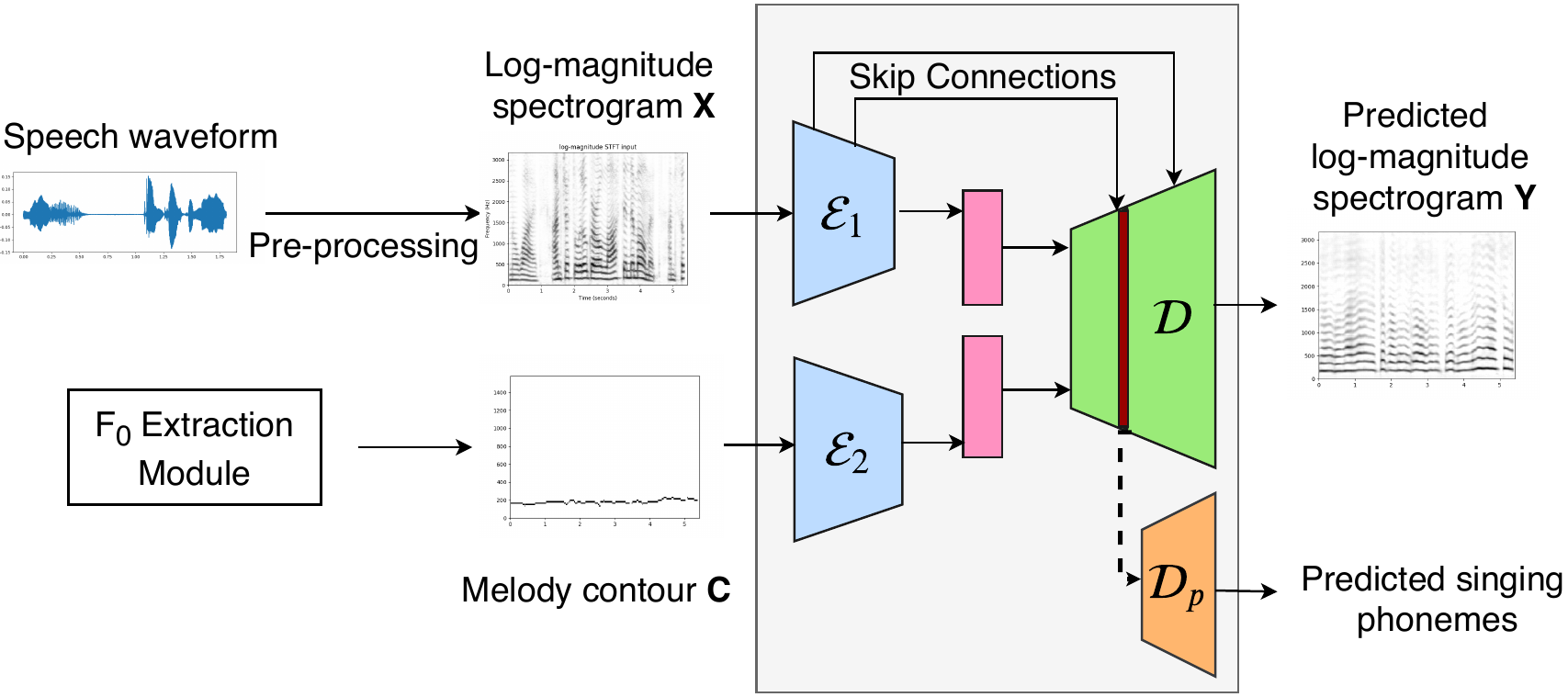}
    \caption[The proposed encoder-decoder framework overview]{Proposed encoder-decoder framework. Given a speech log-magnitude spectrogram $\mathbf{X}$ and melody contour representation $\mathbf{C}$, the encoder--decoder network learns to transform speech into singing and simultaneously predict phonemes in the output singing.}
    \label{main_system_figure}
\end{figure}

\section{Proposed framework}
\label{spsi-pfd}
We adopt a deep learning framework to tackle this problem. Specifically, we perform spectra-to-spectra conversion in an encoder-decoder framework as illustrated in Fig. \ref{main_system_figure}. Herein, we represent the input speech by its log-magnitude spectrogram. A vocal melody extractor \cite{fsu}, is used to extract melody contour from a different source such as humming or reference singing (if available). It is worth noting that since only the $F_0$ contour is required by our system, even if only the musical score is available, the vocal melody contour can be generated using control models proposed by \cite{gotosp2si} for model-based STS. As indicated by our use of the phrase \textit{spectra-to-spectra}, the network is  trained to predict singing log-magnitude spectrogram. Additionally, we predict phonemes in the output singing to improve intelligibility. The whole system is trained through a multi-task learning (MTL) based loss. We now explain each of the components in more detail:

\subsection{Input pre-processing}
\label{preprocess}
Given an input time-domain speech signal and a  target melody/$F_0$ contour the pre-processing consists of the following steps:

\textbf{Silent frame removal}. All the silent frames from the speech signal are removed. This step makes it easier for the network to learn the alignment between speech and singing during training. This is achieved via a short-time energy threshold set at 40 dB below the maximum energy frame. Any set of three or more consecutive silent frames (i.e., longer than approximately 50 ms) are removed.
    
\textbf{Time stretching}. The speech signal and $F_0$ contour can be of widely differing lengths. To make the data suitable for CNN based networks with paired training mechanisms, we time-stretch the input speech to the same length as the target F0 contour using a phase vocoder \cite{librosa}. Here, the time-scaling is controlled by a single factor applied uniformly across the signal (i.e. all the phones). Thus, we see that this step does not use any phoneme alignment information but only the overall duration of target melody contour.
    
\textbf{Log-magnitude representation}. After computing the magnitude spectrogram for the speech signal, we apply an element-wise transformation given by $f(x) = \log(1 + x)$ to obtain a \textit{log-magnitude spectrogram} given by $\mathbf{X} \in \mathbb{R}^{F \times T}$ with $F$ frequency bins and $T$ temporal frames. It is often preferred over spectrograms for input representations as it corresponds better to human perception of sound intensity \cite{goldstein67}.
    
\textbf{Vocal pitch contour}. The input melody/pitch contour is represented in a time-frequency binary image with same resolution and size as that of spectrogram, which is an array of zeros and with a 1 in each time-frame at the nearest frequency index where the extracted melody is present. This is denoted by $\mathbf{C} \in \{0,1\}^{F \times T}$.

\subsection{Network architecture}
We adopt an encoder-decoder based deep learning framework.
We wish to produce two encodings, one for speech and another for the target melody. Moreover, our goal is to use these encodings in conjunction to output a sung version of the speech using a decoder. To do so, we appropriately adapt an encoder-decoder network based on popular U-net \cite{unet} architecture which has also been previously used for singing voice conversion \cite{eric_sing}.
We now discuss some notable features as well as our key modifications for this network over U-net and its version used in \cite{eric_sing}:

\begin{itemize}
    \item Since the network is supposed to handle variable length speech signals, we opt for a fully-convolutional architecture \cite{liu16mm}. Following \cite{eric_sing,liu19ijcai}, we use ``1D convolutional'' \cite{nam19spm} layers rather than 2D convolutional layers, to add flexibility of using recurrent layers in conjunction with the convolutional layers. All recurrent layers in the proposed network are gated recurrent units (GRU) \cite{Chung2014}.
    
    \item The network follows alternating time and frequency downsampling (upsampling) architecture in the encoder (decoder). The downsampling layers use 1D convolutions and the upsampling layers use transposed 1D convolutions. Both the time and frequency axes are downsampled by a factor of 8 to obtain the latent code. We employ skip connections \cite{unet} between encoder $\mathcal{E}_1$ and decoder $\mathcal{D}$, which helps in controlling the gradient vanishing problem and eases training of deeper networks. The encoders $\mathcal{E}_1, \mathcal{E}_2$ have the same architecture and a concatenated version of their produced encoding is served as an input to $\mathcal{D}$.
    
    \item We use instance normalization (IN) layers \cite{instancenorm} before the recurrent GRU layers.\footnote{In our pilot study, we found that using the instance normalization layers before every downsampling/upsampling layer did not offer a significant boost in performance. After experimentation with its placement we decided to use instance normalization before the recurrent GRU layers.} This was primarily motivated due to our belief that normalized input might be more suitable for the GRU layers than input with range $[0, \infty)$ (ReLU activation output) as its range is more in agreement with the range in which the \emph{tanh} activation function operates.
\end{itemize}

\textbf{Phoneme decoder}. When trained with the multi-task learning objective (discussed in Section \ref{training}), we employ a phoneme decoder $\mathcal{D}_{p}$ to predict unnormalized phoneme probabilities $\hat{y}_t^{p} \in \mathbb{R}^{N_p}$ for each time frame $t$ in the output singing. Here $N_p$ is the size of phoneme dictionary, $P=\{0, 1, ..., N_p-1\}$. The input to $\mathcal{D}_{p}$ is the output of the penultimate frequency upsampling block of $\mathcal{D}$. The architecture of $\mathcal{D}_{p}$ consists of a transposed 1D convolutional layer, followed by a GRU and 1D convolutional layer. It is trained jointly with the rest of the network using the frame-level phone labels already available in the dataset.

\subsection{Training}
\label{training}

The training is performed using a multi-task learning (MTL) based objective which consists of two components, 1) mean squared error (MSE) between predicted and true log-magnitude spectrograms $\mathbf{\hat{Y}}, \mathbf{Y} \in \mathbb{R}^{F \times T}$, and 2) average cross entropy (CE) loss for frame-wise phoneme prediction, weighted with hyperparameter $\lambda$. Formally, given an input log-magnitude spectrogram $\mathbf{X}$, melody contour $\mathbf{C}$, predicted unnormalized phoneme probabilities for frame $t$, $\hat{y}_t^{p} \in \mathbb{R}^{N_p}$ (i.e., output of phoneme decoder $\mathcal{D}_p$), and corresponding true phoneme $c_t \in P$ (phoneme dictionary), the following function, averaged over the batch, is optimized during training with respect to $\mathcal{E}_1, \mathcal{E}_2, \mathcal{D}, \mathcal{D}_p$:
\begin{equation*}
        \begin{split}
        & \mathcal{L}_{\text{MTL}} = \mathcal{L}_{\text{MSE}}(\mathbf{Y}, \mathbf{\hat{Y}}) + \frac{\lambda}{T} \sum\limits_{t=1}^T \mathcal{L}_{\text{CE}}(\hat{y}_t^{p}, c_t) \,, \\
        & \mathcal{L}_{\text{MSE}}(\mathbf{Y}, \mathbf{\hat{Y}}) = ||\mathbf{Y} - \mathcal{D}(\mathcal{E}_1(\mathbf{X}), \mathcal{E}_2(\mathbf{C}))||^2 \,, \\
        & \mathcal{L}_{\text{CE}}(\hat{y}_t^p, c_t) = -\hat{y}_t^p(c_t) + \log\big(\sum_{m \in P} \exp(\hat{y}_t^p(m)\big) \,.
        \end{split}
\end{equation*}

\textbf{Data augmentation}. During training, we 
augment the data by pitch-shifting the entire input speech by fixed amount while keeping the target singing spectrogram unchanged. The amount of pitch shift is sampled uniformly at random from the interval $[-1, 1]$ semi-tones.

\subsection{Prediction}

We first recover the magnitude spectrogram from the log-magnitude spectrogram by applying $g(x) = e^x - 1$ element-wise. We then use the popular Griffin-Lim algorithm \cite{griffin-lim}, with a modification to estimate the phase \cite{tacotron}. Herein, we raise the predicted magnitudes in the spectrogram by a power of $1.2$ before applying Griffin-Lim. This reduces artifacts in the final time domain predictions \cite{tacotron}. 

\section{Experimental Validation}
\label{spsi-ev}
\subsection{Setup}

\textbf{Test and Training sample generation}. 
We use the NUS-48E database \cite{nus48e} which consists of 48 recordings of 20 unique English songs, the lyrics of which are sung and read out by 12 subjects (4 recordings each). It contains 115 min of singing data and 54 min of speech data. Each song and speech file is also accompanied by phone-level annotations in accordance to CMU phoneme dictionary (39 phonemes) with two extra phones denoting the silence and inhalation between words. The annotations specify the start and end time of each phone (25,500 phone instances across the dataset). Out of the 20 unique songs in the dataset, we keep one song (with two recordings) as our test set and the others for training. It is important to note that size of the dataset is relatively small. While we attempt to alleviate this problem using data augmentation, given the limited data it is difficult to expect generalizability across different datasets with a variety of singer timbres and lyrics. Thus, we test our system on data which contains singers with minimal presence in training (1 recording each) and on a song \textit{not} seen in training.

We extract our training data from the pre-decided set of training singers/speakers and songs in the following manner: Using the phone-level annotation, first we extract the segments of song that are sandwiched between 2 silences. Most of these silences are more than 200 ms. From these segments, multiple combinations of consecutive words are generated and audio samples corresponding to these sets of words are extracted from both the speech and sung files. All the combinations are constrained to have three or more words. The melody contour is extracted from the sung audio using the system proposed in \cite{fsu}. This set forms our training data. The reasons for adopting such a scheme for training sample generation was to: 1) condition the network better from limited amount of data, and 2) to avoid silences in singing which are often long in duration and make it more difficult to associate input and target spectras during training.

\textbf{Phoneme synchronization (PhSync)}.
To quantify how much the burden of modelling phoneme durations affects the system, we also train our proposed system in a modified setting, which does not require any phoneme duration prediction. We use phase vocoder to stretch/shrink each phone in the input speech to be the same length as it is in the target singing. The duration for each phoneme is obtained from the phone-level annotations for speech and singing. Note that we do not need to explicitly remove silent frames from speech as they are automatically stretched to very small durations in this step.

\textbf{Parameter details}. The time domain signals are resampled from 44k to 16k Hz. We compute STFT with 1024-pt FFT size, 64 ms window size, 16 ms hop size and use the same phoneme dictionary as employed in the dataset. We set $\lambda = 0.015$ for MTL loss and use Adam \cite{adam} as our optimizer with initial learning rate 0.002 and exponential decrease factor of 0.92. The networks are trained for 14 epochs (1000 iterations each) with a batch size of 16. All the neural network implementations and audio processing procedures are performed using PyTorch \cite{pytorch} and librosa \cite{librosa}.

\textbf{Systems evaluated}. \label{systems_evaluated}
As stated before, to our best knowledge there are no machine learning based methods that tackle our task, namely STS conversion without using any additional input such as singing templates. Thus, we create several variants of our system to extensively evaluate the effect of several modifications.
\begin{itemize}
    \item \textbf{Baseline 1 (B1)}: Uses the proposed network with MSE loss ($\lambda=0$) and attempts to directly transform stretched speech to singing with \textit{no} melody information. 

    \item \textbf{Baseline 2 (B2)}: Uses modified version of the proposed network by excluding IN layers and skip connections. Trained with MSE loss.
    
    \item \textbf{AllNorm}: Uses proposed network with IN layers before all upsamling and downsampling layers. Trained with MSE loss.
    
    \item \textbf{Proposed MSE (P--MSE)}: The proposed network (two IN layers before GRU) trained with MSE loss. 
    
    \item \textbf{Proposed MTL (P--MTL)}: This system denotes our proposed network trained with MTL loss ($\lambda > 0$).
    
    \item \textbf{Proposed MSE + PhSync}: Denotes our proposed system trained in phoneme synchronization setting with MSE loss.
    
    \item \textbf{Singing Autoencoder}: This denotes our proposed network (latent code 64 times smaller than input) trained as an autoencoder (with MSE loss) on the NUS-48E singing data. It provides a very strong benchmark to compare the proposed systems against as it reconstructs the input almost perfectly. Its performance can be seen as an upper bound to what can be attained by our systems using the same network architecture.
\end{itemize}

\textbf{Objective evaluation}. We use log-spectral distance (LSD) and $F_0$ raw chroma accuracy (RCA) \cite{mir_eval} to objectively evaluate systems. LSD is used as a metric for phone intelligibility while RCA is used to evaluate melody transfer. LSD is computed by averaging the euclidean distance between true and predicted log-spectrogram frames over time for frequencies between 100 Hz to 3.5 kHz. RCA is computed according to \cite{mir_eval} on pitch contours extracted from the true singing using \cite{fsu}, which is the original contour input to the system and predicted singing using pYIN \cite{pyin}. The systems are evaluated by first selecting random 100 test samples with speech duration of at least 1 second and then averaging the metrics over all the samples.

\textbf{Subjective evaluation}: We conducted preference listening test with 11 participants with normal hearing abilities for a subset of our systems (B2, P--MSE and P--MTL). Each participant was first familiarized with the input speech, target singing and asked to compare the outputs of a selected pair of systems for each of 5 distinct input speech segments (picked randomly from 12 test samples with at least 1 second of speech). For each sample the two systems were randomly chosen. The participants were asked to specify their preference among the two systems (with an additional option of `Tie') for 4 different attributes: Lyrics/Phoneme intelligibility (Q1), Naturalness (Q2), Melodic similarity to target (Q3) and Speaker identifiability (Q4). We report the fraction of votes received by each option for all pairs of systems, over each attribute.

Code, sample outputs and additional material are available at \url{https://jayneelparekh.github.io/icassp20/}.

\subsection{Results and Discussion}

\textit{Objective evaluation}: 
Several observations can be made from objective evaluation results presented in Table \ref{results}.  First, as expected, B1 performs much worse than the other systems, highlighting the importance of using melody for this task. Second, P--MSE performs noticeably better than AllNorm (IN before every layer). This indicates that selectively placing the  normalization layers can improve performance. The system also performs better in the phoneme synchronized setting as it is not required to model any phoneme durations (P--MSE vs Proposed MSE + PhSync). Among the systems using only input speech and target melody for STS conversion, P--MTL performs the best, emphasizing the utility of incorporating singing phoneme prediction loss in our training objective for improving intelligibility. Another notable observation is the comparable and satisfactory melody transfer (as evaluated via RCA) achieved by all systems using the target melody in input.

\begin{figure} [h!]
    \centering
    \includegraphics[width=0.75\columnwidth]{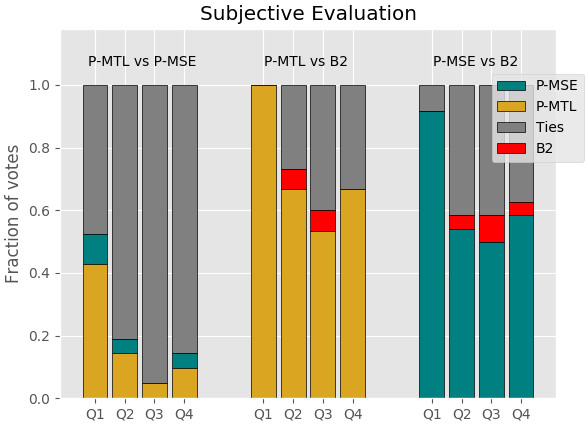}
    \caption[Sample predictions]{Subjective evaluation results for various pairs of systems. Q1: Lyrics/phoneme intelligibility, Q2: Naturalness, Q3: Melodic similarity, Q4: Speaker identifiability are the four attributes.}
    \label{sub_eval}
\end{figure}

\textit{Subjective evaluation}: The listening test results appear in Fig. \ref{sub_eval}. These further validate results on objective metrics. P--MSE and P--MTL were consistently preferred over B2 for all the attributes. As noted before, since all the selected systems perform well in melody transfer, it is the attribute (Q3) that has the largest fraction of tied votes. P--MTL displays better phoneme intelligibility (Q1) than P--MSE but is very similar to P--MSE on other attributes. This further supports our use of MTL loss for better phoneme intelligibility. 

\begin{figure} [h!]
    \centering
    \includegraphics[width=0.91\columnwidth]{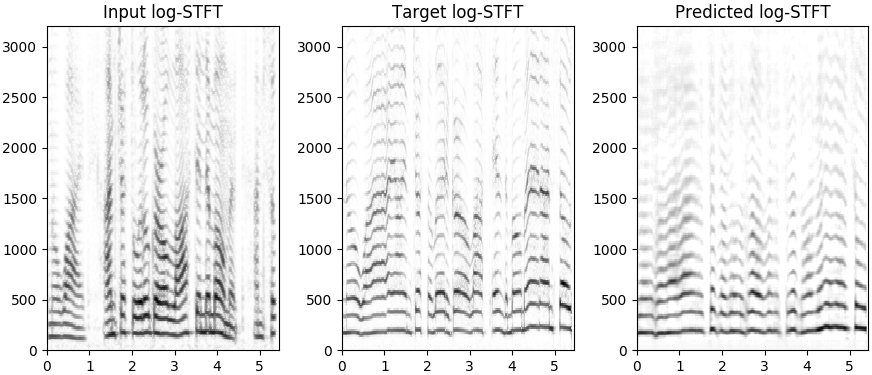}
    \caption[Sample predictions]{Log-magnitude spectrograms for left: time-stretched speech, middle: target singing and right: predicted singing}
    \label{example_prediction}
\end{figure}

\textit{Qualitative observations \& future directions}: 
For our best performing system, P--MTL, we observe that it exhibits promising melody transfer (see Fig. \ref{example_prediction}), fair naturalness of voice, reasonable intelligibility, and models phoneme durations in a plausible manner. The aspect primarily requiring more improvement is speaker identifiability. Employing generative adversarial network (GAN) \cite{gansynth,liu19arxiv} or WaveRNN-like models \cite{wavernn}, exploring use of audio representations like Mel-spectrograms or CQT, or conditioning the network on speaker's \textit{i}-vector \cite{sts_spectral} offer interesting options for future directions to enhance speaker identifiability, and further improve other aspects.

\begin{table}
\centering
\begin{tabular}{l c c c} 
\toprule

System & LSD (dB) $\downarrow$ & RCA $\uparrow$\\
\toprule
Baseline 1 (B1) & 14.19 & 0.221\\
Baseline 2 (B2) & 11.71 & 0.769\\
AllNorm & 11.72 & 0.771\\
Proposed MSE (P--MSE) & 11.22 & 0.829\\
Proposed MTL (P--MTL) & \textbf{10.97} & \textbf{0.857}\\
\midrule
Proposed MSE + PhSync & 10.91 & 0.833\\
\midrule
Singing Autoencoder & 5.51 & 0.991\\
\bottomrule
\end{tabular}
\caption[Results]{
Results on objective metrics for different proposed STS systems (cf. Section \ref{systems_evaluated}). 
Log-spectral distance (LSD) is the lower the better, while raw chroma accuracy (RCA) the inverse. We highlight the best result obtained by the system without phoneme synchronization (PhSync). The last row presents the result of a system that is \emph{not} doing STS. The second last row uses oracle information.
}
\label{results}
\end{table}

\section{Conclusion}
To summarize, we presented the first machine learning based STS conversion system that only requires speech and target melody. It processes their time-frequency representations for generating a ``sung speech." We carefully modify different aspects of the network to improve our predictions. Furthermore, a novel MTL-based loss function is proposed for enhancing phoneme intelligibility. These refinements are validated via both objective and subjective evaluations. Several interesting future directions are also discussed. We hope that our formulation, approach and insights would inspire further research on this interesting topic.

\section{Acknowledgements}
We thank Prof. Preethi Jyothi for her suggestions.
\bibliographystyle{IEEEbib}
\bibliography{strings,refs}

\end{document}